\documentclass[%
reprint,
superscriptaddress,
 amsmath,amssymb,
 aps,prd,
floatfix,
]{revtex4-2}
\usepackage[dvipsnames]{xcolor}

\usepackage{float}

\usepackage{lipsum}
\usepackage{hyperref}
\usepackage{amsmath}
\usepackage{graphicx}
\usepackage{dcolumn}
\usepackage{bm}
\hypersetup{
	colorlinks=true,       
	linkcolor=blue,          
	citecolor=blue,        
	urlcolor=blue           
}

\usepackage{enumitem}
\newlist{STEP}{enumerate}{1}
\setlist[STEP]{label=\Roman*:}

\begin{document}
\title{An exact model for evaporating primordial black holes in cosmological space-time}

\author{Semin Xavier}
    \email{seminxavier@iitb.ac.in}
    \affiliation{Department of Physics, Indian Institute of Technology Bombay, Mumbai, Maharashtra 400076}

\author{Alan Sunny}%
 \email{alansunny21.phy@gmail.com}
\affiliation{Department of Physics, Central University of Tamil Nadu, Thiruvarur, Tamil Nadu 610005 }

\author{S. Shankaranarayanan}
	\email{shanki@phy.iitb.ac.in}
    \affiliation{Department of Physics, Indian Institute of Technology Bombay, Mumbai, Maharashtra 400076}

\date{\today}

\begin{abstract}

Primordial black holes (PBHs) in the mass range $10^{17} - 10^{23}~{\rm gm}$ are considered as possible dark matter candidates as they are not subject to big-bang nucleosynthesis constraints and behave like cold dark matter. If PBHs are indeed dark matter, they cannot be treated as isolated objects in asymptotic flat space-time. Furthermore, when compared to stellar-mass black holes, the rate at which the Hawking particles radiate out from PBHs is significantly faster.  In this work, we obtain an exact time-dependent solution that models evaporating black holes in the cosmological background. As a result, the solution considers all three aspects of PBHs --- mass-loss due to Hawking radiation, black hole surrounded by mass distribution, and cosmological background. Furthermore, our model predicts that the decay of PBHs occurs faster for larger masses; however, \emph{the decay rate reduces for lower mass}. Finally, we discuss the implications of theoretical constraints on PBHs as dark matter.
\end{abstract}
\maketitle


\section{Introduction}

Black holes began as solely a mathematical concept. However, they are currently at the heart of much of high-energy astrophysics, gravitational-wave astronomy, and 
dark matter research~\cite{2019-EventHorizonTelescope,2019-LIGOScientific,LIGOScientific-2021usb,2020-Green-Bradley,2021-Carr_Kuhnel-Astph}. 
Naturally, black holes have been studied extensively in both classical and quantum frameworks~\cite{1998-Frolov.Novikov-Book,2014-Calmet_Bernard,1999-Kokkotas,2008-Psaltis,2009-Sathyaprakash,2014-Clifford,2019-Cardoso_Paolo,2012-Piotr_Lopes}. However, the majority of these studies have concentrated on isolated black holes with two essential properties: the presence of a time-like Killing vector and asymptotic flatness~\cite{2004-Ashtekar_Badri,2008-Emparan_Harvey-LivRRel,2012-Piotr_Lopes,2017-Frolov_David-LivRRel} .

The realistic black holes are embedded in an expanding universe rather than an asymptotically flat (or de Sitter) space-time and are surrounded by local mass distributions rather than being in a vacuum \cite{2000-Nayak_Vishveshwara-PRD}. Thus, while there are uniqueness theorems for stationary black holes, there are no uniqueness theorems for realistic black holes~\cite{1996-Heusler-Book}.  This is especially pertinent for two physically important situations --- black holes evaporating via Hawking radiation and primordial black holes (PBHs). 

Hawking predicted that when quantum matter effects are taken into account, a stationary black hole emits thermal radiation with the Planckian power spectrum characteristic of a perfect black-body at a fixed temperature. However, a radiating black hole is non-stationary as it loses energy and the horizon continuously shrinks. Attempts have been made to model this process using Vaidya-type
metric~\cite{1980-Hajicek_Israel-PhyLetA,1981-Bardeen-PRL,1984-Christensen-Essay,1984-Kuroda-PTP,1985-RLMallett-PRD,1982-Balbinot_Bergamini-NCB,1990-Biernacki-PRD,2005-Nielsen_M.Visser-CQG,2014-M.Visser-PRD,2021-Piesnack_Kassner-grqc}. Metrics of this kind have the advantage of allowing a study of the dynamical evolution of the (apparent) horizons associated with a radiating black hole. However, these models break down at late times because the Hawking temperature increase with loss of mass. In other words, the rate at which the energy is radiated from the black hole also increases.

PBHs are hypothetical black holes that could have formed in the early Universe before the big-bang nucleosynthesis~\cite{2021-Carr_Kuhnel-Astph,1979-Zeldovich_Novikov-NASA,1971-Hawking-MNRAS,1975-Chapline-Nature}. Hence, PBHs are not subject to the well-known big bang nucleosynthesis (BBN) constraint of baryons and can be classified as non-baryonic and behave like any other form of cold dark matter~\footnote{PBH can also form during a matter-dominated epoch that includes deviations from spherical configurations~\cite{2016-Harada_Jhingan-PTEP}}. Furthermore, unlike stellar black holes formed from the collapse of a massive star, PBHs could be produced in the mass range $10^{15} - 10^{30}~{\rm gm}$. Interest in PBH as a dark matter over particle dark matter candidates is because its existence rests on known physics --- general relativity and the presence of primordial fluctuations --- and is independent of the mechanism that generates them~\cite{2021-Carr_Kuhnel-Astph,2021-Menaetal-Frontiers,2018-Sasaki.etal-CQG}. 

Current constraints suggest the PBHs in mass windows $10^{17} - 10^{23}~{\rm gm}$ are potential dark matter candidates~\cite{2020-Green-Bradley,2021-Carr_Kuhnel-Astph}. As mentioned earlier, compared to stellar-mass black holes, the rate at which (Hawking) particles are radiated from PBHs is substantially higher. These have potential observational signatures in gravitational waves and electromagnetic wavebands~\cite{2020-Aggarwal.etal-Arx}. It is then imperative to have a model that describes the evolution of PBHs in the cosmological background and surrounded by mass distributions. To our knowledge, \emph{no exact time-dependent solution} has been found in the literature. 

Sultana-Dyer obtained an exact spherically symmetric black hole solution in expanding cosmological space-time~\cite{2005-Sultana_Dyer}, sourced by non-interacting null dust and normal dust. In this work, we obtain an exact dynamical black-hole space-time in general relativity, which models the evaporation process of such black-holes. We show that the dynamical black-hole has two apparent horizons --- cosmological and dynamical (black hole) horizon. Furthermore, we show that the decay rate of black holes in the cosmological background is \emph{opposite} compared to the black holes in asymptotically flat space-time. Also, the decay of cosmological black holes is faster compared to the Schwarzschild black hole. We discuss the implications of the results for the PBH as dark matter candidates.

We use $(+,-,-,-)$ signature for the 4-D space-time metric. Greek alphabets denote the 4-D space-time coordinates and lower case Latin for radial-temporal plane. We set $8 \pi G = c = 1$. Prime denotes derivative w.r.t $\eta$.

\section{Model and exact solution}

We consider the following action:
\begin{align}
\label{eq:action}
S  = \int d^4x \sqrt{-g} \left[\frac{R}{2}  + L_{\rm fluids} \right] \, ,
\end{align}
where $L_{\rm fluids}$ refers to the Lagrangian density of the non-interacting perfect fluids~\cite{1997-Bicak-PRD,2019-Cote_Faraoni-EUPJ}. 
We consider the following general form to obtain an exact time-dependent spherically symmetric black-hole space-time that models black hole evaporation in cosmological space-time: 
\begin{equation}
\label{eq:SpherMetric}
ds^2 = g_{\alpha\beta} dx^{\alpha} dx^{\beta} = \gamma_{ij} dx^{i} dx^{j} - \rho^2(x^i) d\Omega^2  \, ,  
\end{equation}
where $x^i$ refers to the radial-temporal plane ($i = 1, 2$), $\rho(x^i)$ is the areal radius of the spherical geometry, and $d\Omega^2$ represents the metric on the unit 2-sphere.

Like Sultana-Dyer \cite{2005-Sultana_Dyer}, we will focus on modeling evaporating black hole in the Einstein-de Sitter universe, which is a flat, matter-only Universe: 
\begin{equation}
\label{eq:EinsdS}
ds^2= a^2(\eta) [d\eta^2-dr^2-r^2 d\Omega^2], {\rm where} ~ 
a(\eta) = \left(\frac{\eta}{\eta_0} \right)^2
\end{equation}
$\eta$ is the conformal time, and $\eta_0$ is an arbitrary constant. To model evaporating black-holes in cosmological space-time, we consider the following time-dependent metric in Schwarzschild coordinates $(\bar{t}, \bar{r})$: 
{\small
\begin{align}
\label{eq:SchTD}
    ds^2  &= \left[1-\frac{2 M(\bar{t})}{\bar{r}}\right] d\bar{t}^2 - a^2(\bar{t}) \left[\frac{d\bar{r}^2}{ \left(1-\frac{2 M(\bar{t})}{\bar{r}}\right)} + \bar{r}^2 d\Omega^2 \right] 
\end{align}
}
\!\! where $M(\bar{t})$ is an unknown function of time and $a(\bar{t})$ is the scale factor. Following Sultana and Dyer, applying the following transformations~\cite{2009-Faraoni}: 
\begin{align}
\label{eq:Transf01}
dt = d\bar{t}+\left(\frac{2 M (\bar{t})}{\bar{r}}\right) \frac{ a(\bar{t}) \,  d\bar{r}}{1-\frac{2M(\bar{t})}{\bar{r}}};~~ dr=d\bar{r}
\end{align}
to the line-element \eqref{eq:SchTD} and 
rescaling $dt = a(\eta) d\eta$, leads to 
the following line-element: 
\begin{align}
ds^2 =&  a^2(\eta) \left[ \left(1-\frac{2 M(\eta)}{r}\right) d\eta^2  -\frac{4 M(\eta)}{r} d \eta\,dr \right. \nonumber \\
& 
\label{eq:TDMetric02} 
\qquad ~~ \left. - \left(1+\frac{2 M(\eta)}{r}\right) dr^2-r^2 d\Omega^2  \right] \, .
\end{align}
By solving Einstein's field equations under the following physical conditions, we obtain the exact form of $M(\eta)$: 
First, in the limit of $M(\eta) = M_0$ the time-dependent solution should reduce to Sultana-Dyer black-hole solution~\cite{2005-Sultana_Dyer}. Hence, we assume  $a(\eta)$ corresponds to flat, matter-only Universe as defined in \eqref{eq:EinsdS}. 
Second, since $M_0$ is positive, we demand that $M(\eta) \geq 0$.
Third, the matter source ($L_{\rm fluids}$) for the above line-element is also a combination of two non-interacting perfect fluids --- null dust and a massive dust \cite{2005-Sultana_Dyer,2021-Faraoni}. However, unlike Sultana-Dyer, the energy density of the fluids changes with time.
Lastly, the time-dependent matter sources do not contribute to the stress-tensor along  the axial $(T{^\phi}{_\phi})$ and polar $(T{^\theta}{_\theta})$ directions, i. e., $T{^\theta}{_\theta} = T{^\phi}{_\phi} = 0$. This is consistent with the findings of Candelas~\cite{1980-Candelas-PRD} who showed that renormalized stress-tensor of the quantum field in the Unruh vacuum vanishes along polar and axial.

Varying the action \eqref{eq:action} w.r.t the metric leads to the following Einstein's equation:
\begin{equation}
\label{eq:EinsEq}
    G{^\alpha}{_\beta} = T{^\alpha}{_\beta} = 
    (\mu_{_D}+ p)u{^\alpha} u{_\beta} 
    - p\delta{^\alpha}{_\beta} +\mu_{_N} k^\alpha k_\beta
\end{equation}
where $p$ and $\mu_{_D}$ are the pressure and density of the perfect fluid and $\mu_{_N}$ is the density of the null-fluid. $u^{\alpha}$ and $k{^\alpha}$ are the four velocity and null vector,  respectively. Imposing the above conditions imply that $p = 0$. Note that the four-velocity of the dust is given by $u{^\alpha} = (u^{0},u^{1},0,0)$ and 
for the null fluid $k{^{\beta}} = (k^{0},k^{1},0,0)$. The 4-velocity of the dust and null-fluid satisfies the following normalization conditions: 
\begin{align}
\label{def:Norma}
g_{\alpha \beta}u^{\alpha}u^{\beta}=1,~~~
g_{\alpha\beta}k^{\alpha}k^{\beta}=0~~~{\rm and}~~~g_{\alpha \beta}u^{\alpha}k^{\beta}= 1.
\end{align}

Solving the Einstein's equations \eqref{eq:EinsEq}, we obtain the following two branches of exact solutions: 
\begin{widetext}
{\small
 \begin{align}
ds_{\rm I}{^2} &=  \left(\frac{\eta}{\eta_0}\right)^4\left[\left(1-\frac{2 M_{\rm I}(\eta)}{r} \right) d\eta^2  - \frac{4M_{\rm I}(\eta)}{r} d\eta dr - \left(1+\frac{2M_{\rm I}(\eta)}{r}\right)dr^2-r^2 d\Omega^2\right] 
\quad~\mbox{where}~M_{\rm I}(\eta) = m \left[
\frac{\eta_{\rm  decay}^7}{\eta^7} - 1 \right] 
\label{eqn:S1}\\
    ds_{\rm II}{^2} &=  \left(\frac{\eta}{\eta_0}\right)^4\left[\left(1-\frac{2M_{\rm II}(\eta)}{r}\right) d\eta^2  - \frac{4M_{\rm II}(\eta)}{r} d\eta dr - \left(1+\frac{2M_{\rm II}(\eta)}{r}\right)dr^2-r^2 d\Omega^2\right]\quad~\mbox{where}~M_{\rm II}(\eta) = m \left[1 - 
\frac{\eta_{\rm  decay}^7}{\eta^7}\right]
    \label{eqn:S2}
\end{align}
}
\end{widetext}
%
%
where $\eta_{\rm decay}$ is a positive constant, which sets the decay rate of the black hole.
This is the first key result of this work regarding which we want to the discuss the following points: First, as mentioned above $M(\eta) \geq 0$ implies that $\eta < \eta_{\rm decay}$ for branch I \eqref{eqn:S1}, and  $\eta > \eta_{\rm decay}$ for branch II \eqref{eqn:S2}. Hence, branch I is physically relevant for PBH. Second, the branch II solution \eqref{eqn:S2} approaches Sultana-Dyer solution in the limit $\eta \to \infty$. In the case of branch I, $\eta$ is bounded from above by $\eta_{\rm decay}$. At $\eta = \eta_{\rm decay}$, $M_{\rm I}(\eta) \to 0$ and the metric corresponds to an exact Einstein-de Sitter Universe. We will discuss more in Sec. \eqref{sec:Comparison}.
Third, the above line-elements (\ref{eqn:S1}, \ref{eqn:S2}) are explicitly time-dependent. 
A fundamental feature of the time-dependent space-times, like (\ref{eqn:S1}, \ref{eqn:S2}), is the lack of any (asymptotically timelike) Killing vector field. In Sec. \eqref{sec:Properties}, we discuss the properties of branch I in detail.

Fourth, we have evaluated various Ricci and Riemann invariants, like $R^{\alpha}_{\beta} R_{\alpha}^{\beta} - R^2/2, R_{\alpha\beta\gamma\delta} R^{\alpha\beta\gamma\delta}$. All these quantities have a singularity at $r = 0$. Thus,  both the branches have space-time singularities. 
Fifth, since the two branches are explicitly time-dependent, unlike event-horizon, it is not possible to define the horizon in the space-times globally~\cite{2009-Szabados-LRR}. 
Apparent horizons are defined quasi-locally and are independent of the global causal structure of
space-time. The apparent horizon is a co-dimension two spatial surface (hence local in time) that contains sufficient information regarding the possible formation of an event horizon in the future. We will discuss in Sec. \eqref{sec:Properties}.

Lastly, to physically understand the relation between $M(\eta)$ and the stress-tensor components ($\mu_{_D}, \mu_{_N}$), we rewrite the Einstein's equations in the following form: 
\begin{align}
    G^{0}_{0} &= \mu_{_D}u^{0}\,u_{0} + \mu_{_N} k^{0}\,k_{0} \label{G00}\\
    G^{0}_{1} &= \mu_{_D}u^{0}\,u_{1} + \mu_{_N} k^{0}\,k_{1} \label{G01}\\
    G^{1}_{0} &= \mu_{_D}u^{1}\,u_{0} + \mu_{_N} k^{1}\,k_{0} \label{G10}\\
    G^{1}_{1} &= \mu_{_D}u^{1}\,u_{1} + \mu_{_N} k^{1}\,k_{1} \label{G11} \, ,
\end{align}
We use the normalization conditions \eqref{def:Norma} of the 4-velocity of the dust and null-fluid, and from these equations, we obtain the following relations: 
\begin{align}
\label{def:mu}
   \mu_{_D}&= G^{0}_{0}+G^{1}_{1} \, , \\
   \label{def:tau}
\mu_{_N} &= \left( G^{0}_{1}\,G^{1}_{0}-G^{0}_{0}\,G^{1}_{1} \right)/{\mu_{_D}}   \, .  
\end{align}
Eq. \eqref{def:mu} leads to the energy density of the dust, while Eq. \eqref{def:tau} leads to a definite expression for the null-fluid. The above expression is more simplified compared to Ref.\cite{2005-Sultana_Dyer}. For more details, see Appendix \ref{App:A}.

\section{Geometrical Properties and Invariant quantities}
\label{sec:Properties}

The line-elements (\ref{eqn:S1}, \ref{eqn:S2}) 
are explicitly time-dependent and hence do not 
posses any asymptotically time-like Killing vector to define a preferred time coordinate. Thus, the definition of the surface gravity of the apparent horizon is also ambiguous. However, Kodama proved the existence of a divergence-free vector field for any time-dependent spherically symmetric metric of the form \eqref{eq:TDMetric02}~\cite{1980-Kodama-PTP,2006-Racz-CQG}. Interestingly, one can use the Kodama vector to obtain an invariant definition of the surface gravity of the apparent horizon~\cite{2009-Criscienzo_Hayward-CQG,2016-Faraoni.etal-PRD}. 

Another invariant quantity that is useful to locate the apparent horizon is the Misner-Sharp-Hernandez energy $(E_{\rm MSH})$~\cite{1970-Cahill.McVittie-JMP,1996-Hayward-PRD,2009-Szabados-LRR}. For the general spherically symmetric line-element \eqref{eq:SpherMetric}, Misner-Sharp-Hernandez energy satisfies the scalar equation $\rho_{\rm AH} = 2 E(\rho_{\rm AH})$  at the apparent horizon (AH). Interestingly, Misner-Sharp-Hernandez energy is the conserved Noether charge corresponding to the conservation of the Kodama current~\cite{1996-Hayward-PRD}. In the rest of this section, we obtain the following invariant quantities that describe the properties of the line-element \eqref{eqn:S1}: Apparent horizon, Misner-Sharp energy, Kodama vector, and the associated current. Since these quantities are invariant, we obtain them w.r.t the conformal time ($\eta$). We also obtain the conditions on the energy density. For branch II, see Appendix \eqref{App:prop}.

\subsection{Apparent horizon}

For the line-element \eqref{eq:SpherMetric}, we can define the following scalar quantity~\cite{1996-Hayward-PRD,2009-Hayward.etal-CQG,2009-Szabados-LRR}:
\begin{equation}
\label{def:chi}
\chi(x) = \gamma^{ij}(x) \partial_i \rho \, \partial_j \rho .  
\end{equation}
The following conditions give the apparent horizon:
\begin{equation}
\label{def:AHcon}
\chi(x )\big|_{\rm AH} = 0 \, ; \quad 
\partial_i \chi(x) \big|_{\rm AH} \neq 0. 
\end{equation}
For the line-element \eqref{eqn:S1}, $\chi(\eta,r)$ is given by:
{\small
\begin{align}
\label{eq:chiS1}
\!\!\!\!\!\! \chi(\eta,r)= (r H - 1) \left[
\left[1 - \frac{2M_{I}(\eta)}{r}\right]  + 
\left[1 + \frac{2M_{I}(\eta)}{r}\right] r H
\right]
\end{align}
}
where $H \equiv H(\eta) = {a}'(\eta)/a(\eta) = 2/\eta$. Using the conditions \eqref{def:AHcon}, the horizons are at:
{\small
\begin{align}
r_{\rm C} & = {1}/{H} \label{eqn:DymHr1}\\
r_{\rm H} & = \left(M_{I}(\eta) + \frac{1}{2H} \right) \left[\sqrt{1 + \frac{8 H M_{I}(\eta)}{(1 + 2 H M_{I}(\eta))^2}} - 1 \right].
    \label{eqn:DymHr2}
\end{align}
}
Note that the apparent horizon \eqref{eqn:DymHr1} is due to cosmological expansion~\cite{2018-Melia-AJP} and is also present in the case of 
Sultana-Dyer~\cite{2005-Sultana_Dyer}. In the limit of $M_I(\eta) \to M_0$, the apparent horizon \eqref{eqn:DymHr2} corresponds to the particles closest to the event horizon crossing the superluminal barrier (cf. Eq. 26 in \cite{2005-Sultana_Dyer}). Thus, the above apparent horizon reduces to the event-horizon for constant mass. At $\eta \to \eta_{\rm decay}$, $r_{\rm H}$ vanishes indicating that the apparent horizon ceases to exist.

The dynamic surface gravity associated with the apparent horizon of the line-element \eqref{eq:SpherMetric} is given by~\cite{2009-Hayward.etal-CQG}:
\begin{equation}
\label{def:SurGrav}
\kappa_{\rm AH}= \left.\frac{1}{2 \sqrt{-\gamma}} \partial_{i}\left(\sqrt{-\gamma} \gamma^{i j} \partial_{j} \rho \right)\right|_{\rm AH}
\end{equation}
Since it is a scalar quantity it is also an invariant quantity. For the line-element \eqref{eqn:S1} and apparent horizon \eqref{eqn:DymHr1}, we have:
\begin{equation}
    \kappa_{\rm C}=\frac{1}{2 \, a(\eta)}\left(\frac{H}{2} - 3M_{\rm I}(\eta)\,H^2\right). 
    \label{eqn:Kappar1}
\end{equation}
In the limit of $\eta \to \eta_{\rm decay}$, $M_{I}(\eta) \to 0$, and the line-element \eqref{eqn:S1} corresponds to that of a pure 
Einstein-de Sitter, i. e.,:
\begin{align}
 \kappa_{\rm C} &= \frac{H}{4\,a(\eta_{\rm decay})} \, \nonumber.
\end{align}
For the line-element \eqref{eqn:S1} and apparent horizon \eqref{eqn:DymHr2}, we have:
{\small
\begin{align}
  \kappa_{\rm H}&= - \frac{1}{a(\eta)} 
  \left[\frac{1}{r_{\rm H}} \left(\frac{M_{\rm I}(\eta)}{r_{\rm H}} + M'_{\rm I}(\eta) \right) 
  + H \left(\frac{M_{\rm I}(\eta)}{ r_{\rm H}} - M'_{\rm I}(\eta)\right) \right. \nonumber \\
& \qquad \qquad \left. - \frac{a''(\eta)}{2 a(\eta)} \left(r_{\rm H} + 2M_{\rm I} (\eta)\right)  
  \right] \, .
\end{align}
}

As noted earlier, in the limit $\eta \to \eta_{\rm decay}$,
the line-element \eqref{eqn:S1} becomes Einstein-de Sitter and $r_{\rm H}$ vanishes. Hence, in this limit $\kappa_{\rm H}$ is not defined. In the limit $\eta  \rightarrow 0$, $\kappa_{\rm H}$ is ill-defined as the metric diverges. At the limit $\eta_{\rm decay} \to 0$ we have:
{\small
\begin{align}
    \kappa_{\rm H} = \left(\frac{1}{a(\eta)}\right) \frac{28m - \eta +\sqrt{16m^2 - 24m\,\eta + \eta^2}}{\left[4m - \eta + \sqrt{16m^2 - 24m\,\eta + \eta^2}\right]^2} \, .
\end{align}
}

\subsection{Misner-Sharp-Hernandez energy} 
\label{sec:MSH}

In terms of $\chi(x)$ and areal radius $\rho$, the Misner-Sharp-Hernandez energy is~\cite{2009-Szabados-LRR}:
\begin{equation}
\label{def:MSH}
E_{\rm MSH}(x) = 
\frac{\rho(x)}{2}  \left(1 - \chi(x)\right).    
\end{equation}
Substituting $\chi$ from Eq. \eqref{eq:chiS1}, we have: 
\begin{align}
E_{\rm MSH}(r,\eta) &=\frac{r}{2} a(\eta) \left[ 1 + (1 - r H) \left( \left[1- \frac{2M_{I}(\eta)}{r}\right] \right. \right. 
\nonumber \\ 
\label{eq:MSH-S1}
 & \qquad \qquad \quad \left. \left.
+ \left[1 + \frac{2M_{I}(\eta)}{r}\right] r H  \right)\right].
\end{align}
The above expression provides some crucial features about the 
physically realizable values $\eta$ can take for the energy to be non-negative. First, substituting $H = 2/\eta$ in the above expression, we see that the Misner-Sharp-Hernandez energy is positive-definite only if $\eta - 2 r > 0$. In the case of $\eta - 2 r < 0$, the second term in the RHS will be negative, and $E_{\rm MSH}(r,\eta)$ is not always positive definite. Second, in the limit $\eta \to \eta_{\rm decay}$, the Misner-Sharp-Hernandez energy reduces to 
\begin{equation}
 E_{\rm MSH}(r,\eta_{\rm decay}) = 
 r \left(\frac{\eta_0}{\eta_{\rm decay}}\right)^2 \left[1 -  2 \frac{r^2}{\eta_{\rm decay}^2}
 \right]. 
 \end{equation}
Here again we notice that, $E_{\rm MSH}$ is positive only if $\eta_{\rm decay} > \sqrt{2} r$. Third, in the limit of $\eta_{\rm decay} \to 0$, the Misner-Sharp-Hernandez energy \eqref{eq:MSH-S1} reduces to:
\begin{equation}
 E_{\rm MSH}(r,\eta) = \frac{r}{2} a(\eta)
 \left[2 - r^2 H^2 - \frac{2 m}{r} (r H - 1)^2 \right] .
 \end{equation}
This again gives a condition on $\eta$ for which energy is positive definite. 

To further investigate this, we now look at the energy density of the dust given in Eq. \eqref{def:mu}. It is important to note that while the Misner-Sharp-Hernandez energy is a scalar quantity, the energy density of the dust is not scalar. However, it provides crucial insights into the energy conditions of the matter fields. Substituting Einstein tensor in Eq. \eqref{def:mu}, we have:
\begin{align}
\label{eq:muS1a}
    \mu_{_D}&= \frac{H^2}{a^2(\eta)}  \left[\Tilde{\sigma}_{D}(r,\eta) +  
    \Tilde{\sigma}_{N}(r, \eta)\,(\eta - r)\right]
    \end{align}
where 
\begin{align}
\Tilde{\sigma}_{D}(r,\eta) = 3 - \frac{7m\,\eta_{\rm decay}^7}{r\,\eta^7} \, ,
\Tilde{\sigma}_{N}(\eta, r) =  \frac{m}{r^2} \left[\frac{\eta_{\rm decay}^7}{\eta^7} + 6\right]
\end{align}
represent the flow of energy along the radial direction and the null direction, respectively. At $\eta \to \eta_{\rm decay}$, when $M_{\rm I} \to 0$ the flow of energy along the radial direction is constant while the flow of energy along the null decays:
\begin{align}
\Tilde{\sigma}_{D}(r,\eta_{\rm decay}) \to 3 - \frac{7m}{r} \, , \Tilde{\sigma}_{N}(\eta_{\rm decay}) \to  \frac{7 m}{r^2} 
\, . 
\end{align}
This implies the following: As the mass of the black hole decreases, the massless radiated (Hawking) particles that emit decrease. Since the metric asymptotes to Einstein-de Sitter, the stress-tensor of the matter asymptotes to a constant value. Rewriting Eq. \eqref{eq:muS1a}, we have:
{\small
\begin{equation}\label{eq:muS1}
    \mu_{_D}=\frac{4 \eta_{0}^{4}}{r^{2} \eta^{6} }\left[
    3r^2-\left[  \frac{8\,\eta_{\rm decay}^7}{\eta^7}+6\,\right]mr +\left[6+\frac{\eta_{\rm decay}^7}{\eta^7}\right]\,m\eta \right] 
\end{equation}
}
The positivity of the energy condition implies
\begin{align}
 3r^2-\left[\frac{8\,\eta_{\rm decay}^7}{\eta^7}+6\,\right] mr +\left[ 6+\frac{\eta_{\rm decay}^7}{\eta^7}\right] m\eta&>0 \, .
\end{align}
Treating this as a quadratic equation in $r$, and using the fact that the above condition corresponds to a parabola that does not intersect the horizontal axis leads to the following condition:
\begin{align}
     \,\left(  \frac{8\,\eta_{\rm decay}^7}{\eta^7}+6\,\right)^2-12\,\left(6+\frac{\eta_{\rm decay}^7}{\eta^7}\right)\,\frac{\eta}{m}<0. \label{energydensitycondition}
\end{align}
It is interesting to note that the above condition is independent of the value of $r$, implying that the condition is valid for all values of $r > r_{\rm H}$. Thus, for example, in the limit, $\eta_{\rm decay} \gg \eta$, the condition \eqref{energydensitycondition} reduces to:
\begin{align}
    \left(16m\,\eta_{\rm decay}^7 /3\right)^{1/8}< \eta < \eta_{\rm decay}. 
\end{align}
In the limit of $\eta \to \eta_{\rm decay}$, the condition \eqref{energydensitycondition} implies:
\begin{align}
m < 3\eta_{\rm decay}/7 .    
\end{align}
This condition also implies at as $\eta_{\rm decay} \to 0$, $m \to 0$ for the energy-density to be positive.

\subsection{Kodama vector and current}
\label{sec:Kodama}

For line-element \eqref{eq:SpherMetric}, the Kodama vector is~\cite{1980-Kodama-PTP,1996-Hayward-PRD,2011-Vanzo.etal-CQG}:
\begin{align}
\label{def:Kodama}
K^i = \epsilon^{ij} \partial_i \rho(x)    
\end{align}
where $\epsilon^{ij}$ is the volume form associated with $\gamma_{ij}$. Note that the Kodama vector lies in the plane orthogonal to the sphere of symmetry; hence, the Kodama vector along $\theta$ and $\phi$ is zero~\cite{1980-Kodama-PTP}. For the line-element \eqref{eqn:S1}, the Kodama vector is:
\begin{align}\label{eq:KodamaVec}
    K^{0}(x)=\frac{1}{a(\eta)} \, , \,   
     K^{1}(x)=- \frac{r H}{a(\eta)}  \, .
\end{align}
From Eq. \eqref{eq:MSH-S1}, it is easy to see that 
the above Kodama vector satisfies the following relation:
\begin{align}
K^2 = \frac{2 E_{\rm MSH}}{\rho} - 1  \, . 
\end{align}
Since the Kodama vector is conserved ($\nabla_{i} K^i = 0$), we can construct an associated current
\begin{equation}
J^i = G^{i}_{j}\,  K^j   \, ,  
\end{equation}
that is also conserved. For the line-element \eqref{eqn:S1}, the associated current is:
\begin{align}
J^0 & = \frac{H^2}{a^3(\eta)}\left(3 + \frac{4M_{I}(\eta)}{r^2} \left[r - \frac{1}{H}\right] \right)\\
J^1 & =  \frac{2 \left[r H - 1 \right]}{a^3(\eta) r^2}  \left[ H M_{I}(\eta) - M'_{I}(\eta)\left(r H - 1\right) \right] \, .
\end{align}
To obtain the conserved charges, we need to fix the 3-space at a fixed time $\eta$. 
 
\section{Comparison with Schwarzschild}
\label{sec:Comparison}

As mentioned in the Introduction, the original derivation of Hawking assumed that the space-time is static or stationary. This assumption is valid only when the radiated energy is negligibly small compared with the mass-energy of the black hole. However, when the radiation becomes sufficiently large, back-reaction effects will modify via the semi-classical Einstein equation~\cite{1994-Wald-Book}. However, this is highly non-trivial for the four-dimensional space-time~\cite{1980-Candelas-PRD}. As shown by Page~\cite{1976-Page-PRD}, if we only include massless fields, the Schwarzschild black-hole mass (${M}_S$) decays as:
\begin{equation}
\label{eq:MdecaySS}
\frac{d M_S(t)}{dt} = - \frac{1}{3} \frac{1}{t_{\rm decay} M_S^2} \,  \longrightarrow \, {\cal M}_S(\tau) =  \left(1 - \tau \right)^{1/3}  \, 
\end{equation}
where $\tau \equiv {t}/{t_{\rm decay}}$ and ${\cal M}_S$ 
is the (dimensionless) rescaled mass. Attempts are made to model the decay using Vaidya-type metric~\cite{1980-Hajicek_Israel-PhyLetA,1981-Bardeen-PRL,1984-Christensen-Essay,1984-Kuroda-PTP,1985-RLMallett-PRD,1982-Balbinot_Bergamini-NCB,1990-Biernacki-PRD,2005-Nielsen_M.Visser-CQG,2014-M.Visser-PRD,2021-Piesnack_Kassner-grqc}. However, all these analyses are restricted to asymptotically flat space-times in a non-expanding space-time. 

\begin{figure*}[!htb]
\begin{minipage}[b]{.5\textwidth}
\includegraphics[width=\columnwidth]{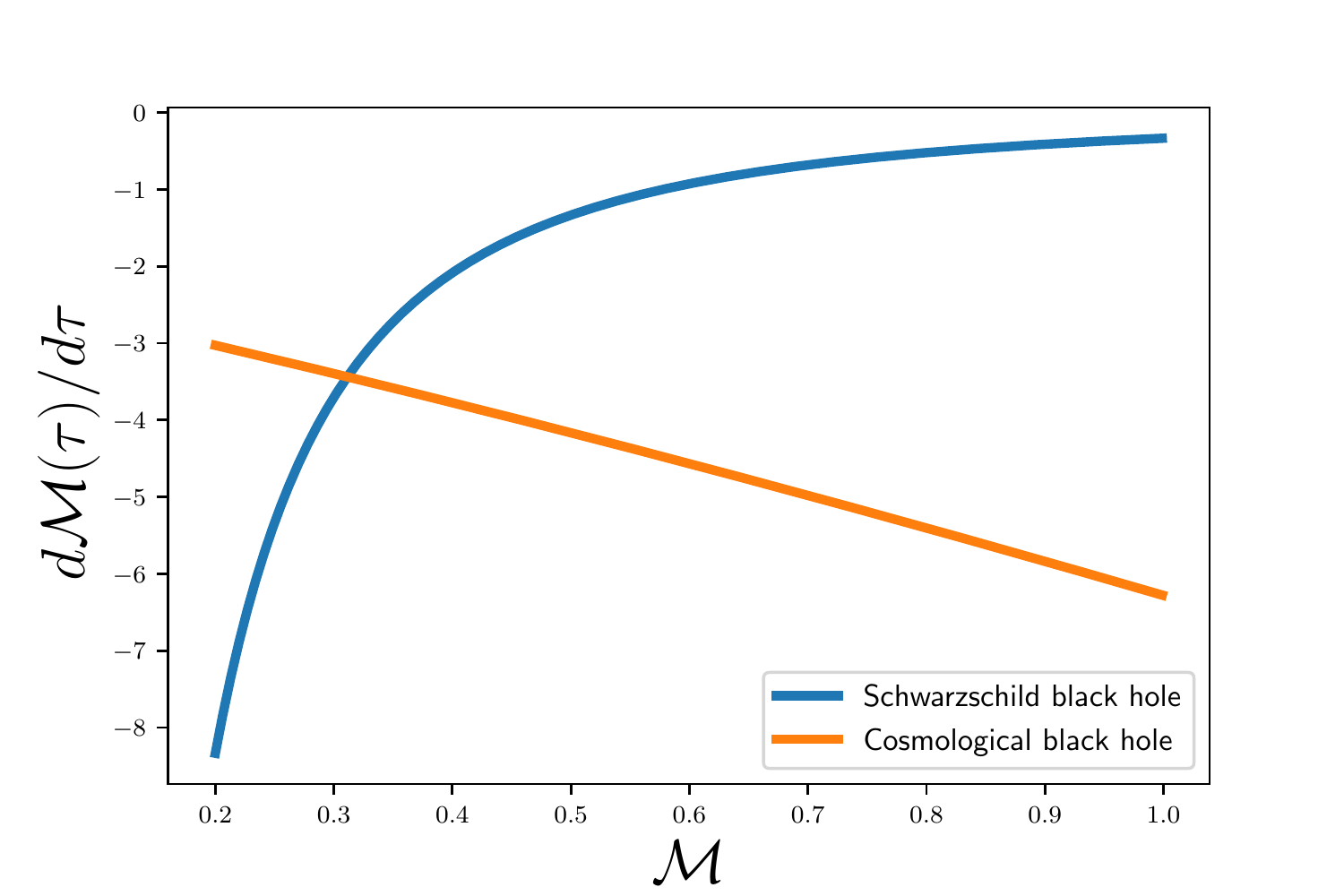}
\end{minipage}\hfill
\begin{minipage}[b]{.5\textwidth}
\includegraphics[width=\columnwidth]{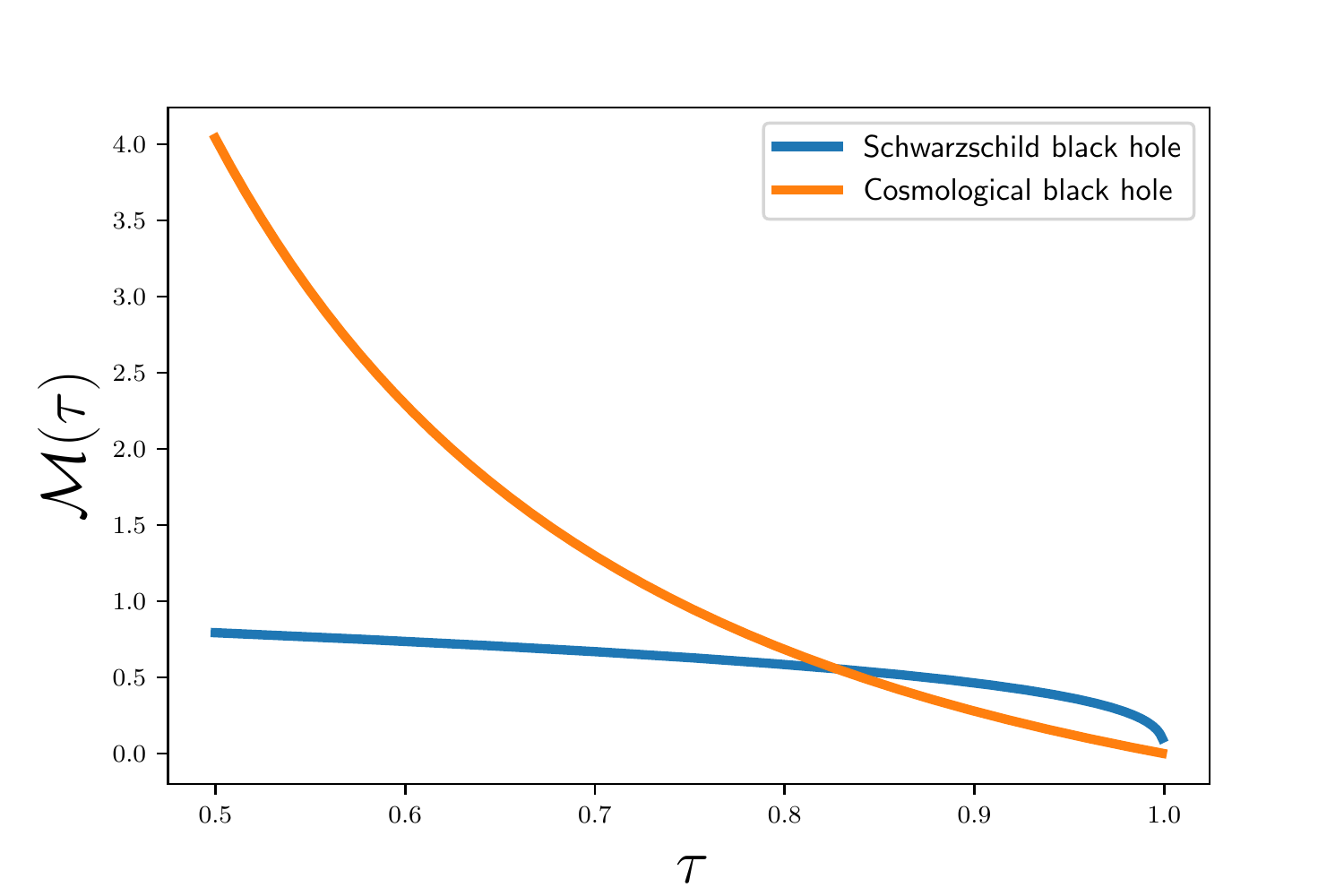}
\end{minipage} 
\caption{Left: Plot of rate of change of mass with time versus ${\cal M}$. Right: Plot of ${\cal M}$ as a function of $\tau$ for Schwarzschild for Schwarzschild \eqref{eq:MdecaySS} and cosmological black hole \eqref{eq:MdecaySD}.}
\label{fig:01}
\end{figure*}

Eq. \eqref{eq:muS1a} implies that the 
space-time asymptotes to Einstein-de Sitter, the stress-tensor of the matter asymptotes to a constant value while the flow of null particles decays. Thus, the line-element \eqref{eqn:S1} models an evaporating black-hole in an expanding FRW background with a matter surrounding it. Thus, the mass of the black-hole in the line-element \eqref{eqn:S1} decays as:
\begin{equation}
\label{eq:MassSD}
{\cal M}_{\rm D}(\tau) =  \left[{\tau}^{-7/3} - 1 \right] 
\end{equation}
where we have relation between the conformal time $(\eta)$ and cosmic time $(t)$, i. e., $t = \eta^3/(3 \eta_0^2)$ and ${\cal M}_D = M_{I}/m$. From the above expression, we obtain the following decay rate for the black-hole in cosmological space-times:
\begin{equation}
\label{eq:MdecaySD}
\frac{d {\cal M}_{\rm D}(\tau)}{d\tau} = - \frac{7}{3} \left({\cal M}_{\rm D} + 1 \right)^{10/7} \, .
\end{equation}
This is another key result of this work regarding which we want to discuss the following points: First, comparing Eqs. \eqref{eq:MdecaySS} and \eqref{eq:MdecaySD}, in the case of Schwarzschild black holes, we see that the decay rate of smaller black-holes is larger compared to solar mass black holes. However, in the case of black holes in cosmological background, it is the opposite; the decay of larger black holes is faster. The same can be seen in the left panel of Fig. \ref{fig:01}. Second, comparing Eqs. \eqref{eq:MdecaySS} and \eqref{eq:MassSD}, we see that the decay of a cosmological black hole is faster compared to the Schwarzschild black hole at the initial phase. However, when the mass is smaller, the Schwarzschild black hole decays faster. See the right panel of Fig.  \ref{fig:01}. Third, since the decay for the smaller black hole is slow in the cosmological background, it is possible to study the end-stages of black hole evaporation in a controlled manner.

These two results have potentially important implications for the primordial black holes as a dark matter candidate. Irrespective of the PBH mass, our model suggests that decay of the primordial black holes occurs faster; however, for lower masses, the decay rate falls. More importantly, it has been argued that the primordial black holes less than $10^{15}~{\rm gm}$ must have evaporated by now~\cite{2021-Carr_Kuhnel-Astph,2021-Menaetal-Frontiers,2018-Sasaki.etal-CQG}. These analyses assume the evolution of an isolated, asymptotically flat space-time. However, this assumption may not be valid for realistic black holes in cosmological space-time and must be reconsidered. In other words, it is not possible to completely rule out primordial black holes less than $10^{15}~{\rm gm}$.

\section{Discussions and Conclusions} 

This work obtains an exact time-dependent black hole solution that models evaporating black holes in the cosmological background. Thus, the solution considers all three aspects of PBHs --- Hawking radiation, black hole surrounded by mass distribution, and cosmological background. We have shown that as the mass of the black hole decreases, the massless radiated (Hawking) particles that emit decrease. Since the metric asymptotes to Einstein-de Sitter, the stress-tensor of the matter asymptotes to a constant value. Our model predicts that decay of the PBHs occurs faster; however, as the mass is reduced, the decay rate reduces. 
We have also discussed the implications of our work regarding the theoretical constraints on primordial dark matter. 

The above result brings attention to the following interesting questions: 
\begin{enumerate}
    \item To exactly quantify the spectrum of Hawking radiation, we need to extend the analysis for the dynamical horizon. This has been discussed in the literature~\cite{2011-Vanzo.etal-CQG}, however, in our case, we do not have well-defined asymptotic vacuum states.
     \item Since the decay rate is slow for smaller black holes, it is possible to study the end-stages of black hole evaporation in a controlled manner than in the case of asymptotically flat space-times. This is currently under investigation.
    \item The analysis rests on spherical symmetry. We need to extend the analysis to axially symmetric space-time. This is currently under investigation.
    \item The constraints on PBHs as dark matter candidates are based on the black-holes in asymptotically flat space-time. Thus, the constraints need to be reworked based on the exact model proposed here with two arbitrary parameters $m$ and $t_{\rm decay}$.
\end{enumerate}

\begin{acknowledgments}
The authors thank Susmita Jana, Ashu Kushwaha, and S. Mahesh Chandran for discussions. SX is financially supported by the MHRD fellowship at IIT Bombay. This work is supported by SERB-MATRICS grant.
\end{acknowledgments}

\appendix

\section{Obtaining $\mu_{_N}$ and $\mu_{_D}$}
\label{App:A}

In this appendix, we show that the solutions (\ref{eqn:S1}, \ref{eqn:S2}) are indeed unique solutions to Einstein's field equations with time-dependent matter consisting of null and massive dust.

For the line-element \eqref{eq:TDMetric02}, we have:
\begin{align}
    G{^\theta}{_\theta} = G{^\phi}{_\phi} = \frac{\eta_{0}^4\left[\ddot{M}(\eta) \eta + 8\dot{M}(\eta)\right]}{\eta^5 r}
\end{align}
Demanding that the fluids are null and massive dust leads to the condition $G^{\theta}_{\theta} = G^{\phi}_{\phi} = 0$:
\begin{align}
    & \frac{\eta_{0}^4\left[\ddot{M}(\eta) \eta + 8\dot{M}(\eta)\right]}{\eta^5 r} =0
\end{align}
Thus, we have: 
\begin{align}
    & M(\eta) = C_{1} + \frac{C_{2}}{\eta^7}.
\end{align}
where $C_{1} = \pm m$ and $C_{2} = \mp m \, \eta_{decay}^7$ are constants. The positive sign in $C_1$ corresponds to branch I \eqref{eqn:S1} and negative sign in $C_1$ corresponds to branch II \eqref{eqn:S2}.

The energy-momentum tensor components for the two branches are:
{\small
\begin{widetext}
\begin{align}\label{SETBranchI}
    T^{a}_{b\,(\text{I})}&=\left[\begin{array}{cccc}
\frac{4 \eta_{0}^{4}\left(-\sigma_{N}\,r - 2\,M_{I}(\eta)\,\eta + 3r^2\right)}{r^{2} \eta^{6}} & \frac{2 \eta_{0}^{4}\left(2\,M_{I}(\eta) - r(3 -\sigma_{D})\right)}{r^{2} \eta^{5}} & 0 & 0 \\
-\frac{2 \eta_{0}^{4}\left(12rM_{I}(\eta) -r\eta\,(3 -\sigma_{D}) + 2\,M_{I}(\eta)\eta\right)}{r^{2} \eta^{6}} & -\frac{4\eta_{0}^{4}\left(4\,M_{I}(\eta)\eta + r(3 -\sigma_{D})[r -\eta]\right)}{r^{2} \eta^{6}} & 0 & 0 \\
0 & 0 & 0 & 0 \\
0 & 0 & 0 & 0
\end{array}\right]
\\
T^{a}_{b \, (\text{II})}&=\left[\begin{array}{cccc}
\frac{4\eta_0^4\left(\tilde{\sigma}_{N}\,r -2M_{II}(\eta)\,t +3r^2\right)}{r^2t^6} & \frac{2\eta_0^4\left((\tilde{\sigma}_{D}-3)r +2M_{II}(\eta)\right)}{r^2\,\eta^5}& 0 & 0 \\
-\frac{2\eta_0^4\left(12M_{II}(\eta)\,r + (\tilde{\sigma}_{D}-3)\,r\eta+2M_{II}(\eta)\,\eta\,\right)}{r^2\eta^6} &  - \frac{4\eta_0^4\left(4M_{II}(\eta)\eta- r\,(\tilde{\sigma}_{D}-3)(r -\eta)\,\right)}{r^2\eta^6} & 0 & 0 \\
0 & 0 & 0 & 0 \\
0 & 0 & 0 & 0
\end{array}\right]
\end{align}
\end{widetext}
where 
\begin{align}\label{SETBranchII}
     \sigma_{D} &= 3 - \frac{7m\,\eta_{decay}^{7}}{r\,\eta^7} \\
      \sigma_{N} &=   m \left(\frac{\eta_{decay}^{7}}{\eta^7} + 6\right) \\
      \tilde{\sigma}_{D} &=  3 + \frac{7m\,\eta_{decay}^7}{\eta^7r} \\
     \tilde{\sigma}_{N} &=  m\left(6 +\frac{\eta_{decay}7}{\eta^7}\right)
\end{align}
}

{
Eqs. (\ref{G00}, \ref{G01}, \ref{G10}), and \eqref{G11} can be written in the following matrix form: 
\begin{equation}
\left[\begin{array}{llll}
 \quad G^{1}_{ 0} & -\,G^{0}_{ 0} & \quad 0 & -\mu_{N}  \\
\quad 0 & -\mu_{D} & \quad G^{1}_{ 0} & G^{1}_{ 1}    \\
-\,G^{1}_{1} & \quad  G^{0}_{ 1} & -\mu_{N} & 0 \\
-\,\mu_{D} & \quad 0 & \quad G^{0}_{0}& G^{0}_{1}
\end{array}\right]\left[\begin{array}{l}
u^{0} \\
u^{1} \\
k^{0} \\
k^{1}
\end{array}\right]=\left[\begin{array}{l}
0 \\
0 \\
0 \\
0
\end{array}\right] \, .
\label{eq:Matrix}
\end{equation}
}
{Using the Gauss elimination method, the above matrix can be written as
{\small
\begin{widetext}
\begin{equation}
\left[\begin{array}{llll}
\quad  G^1_0 & - G^0_0 & \quad 0 & -\mu_{_N}  \\
\quad 0 & -\mu_{_D}& \quad  G^1_0 & \quad  G^1_1    \\
\quad 0 & \quad 0 & \frac{\left( G^1_0\, G^0_1- G^0_0\, G^1_1-\mu_{_D}\, \mu_{_N}\right)}{\mu_{_D}}& \frac{G^1_1\,\left( G^1_0\, G^0_1- G^0_0\, G^1_1-\mu_{_D}\, \mu_{_N}\right)}{\mu_{_D} \,G^1_0} \\
\quad 0 & \quad 0 &  \quad 0 & \frac{\left( G^1_0\, G^0_1- G^0_0\, G^1_1-\mu_{_D}\, \mu_{_N}\right)}{G^1_0}
\end{array}\right]\left[\begin{array}{l}
u^{0} \\
u^{1} \\
k^{0} \\
k^{1}
\end{array}\right]=\left[\begin{array}{l}
0 \\
0 \\
0 \\
0
\end{array}\right]\,.
\label{eq:GaussEli-metric}
\end{equation}
 \end{widetext}
 }
 }
{We obtain non-trivial solutions for the massive dust $(u^0, u^1)$ and the null dust $(k^0, k^1)$, using the following three steps:}
\begin{STEP}
\item  {\emph{To obtain $\mu_N$ and $\mu_D$.} One possible choice is to set 
\begin{equation}
    G^{1}_{0}\, G^{0}_{1}-G^{0}_{0}\, G^{1}_{1}=\mu_{D}\mu_{N}.
    \label{eq:condition01}
\end{equation}
The above condition leads to the fact that $k^{1}$ is arbitrary and non-vanishing. Using Eq.~(\ref{def:mu}) which gives $\mu_{D}$, we can determine $\mu_{N}$ from the above expression. Note that the above conditions are also satisfied for the Sultana-Dyer metric~\cite{2005-Sultana_Dyer}.}

{The third row of the matrix \eqref{eq:GaussEli-metric} is:
{\small
\begin{equation}
   \left( \frac{G^{1}_{0}G^{0}_{1}-G^{1}_{1}G^{0}_{0}-\mu_{D}\mu_{N}}{\mu_D}\right)\left( k^0 +\frac{G^{1}_{1}}{G^{1}_{0}}\,k^1 \right)=0
\end{equation}}
}
{In order for $k^1$ to be unique (apart from the normalising condition $g_{ab}k^{a}\,k^{b}=0$), $k^0$ and $k^1$ must satisfy the following condition:
\begin{equation}
   \left( k^0 +\frac{G^{1}_{1}}{G^{1}_{0}}\,k^1\right)\neq 0 
\end{equation}
In the case of Sultana-Dyer metric, since $k^1 =-k^0$ and $G^{1}_{1}\neq G^{1}_{0} $, the above condition is automatically satisfied for Sultana-Dyer metric. In this step, we have obtained explicit expressions for $\mu_N$ and $\mu_D$ and condition on $k^0$ and $k^1$.
}
\item { \emph{To obtain explicit relations for $(u^0, u^1)$ in-terms of $(k^0, k^1)$.} To go about this, we consider the first and second rows of Eq. \eqref{eq:Matrix}:
\begin{align}
   G^{1}_{0} \,u^0 -G^{0}_{0}\, u^1&=\mu_{N} \,k^1\\
     G^{1}_{0} \,k^0 +G^{1}_{1}\, k^1&=\mu_{D}\,u^1 
\end{align}
Rewriting these expressions, we have:
\begin{align}
   G^{1}_{0} \,u^0 -\frac{G^{0}_{0}}{\mu_{D}} \mu_{D} u^1=\mu_{N} \,k^1  \\ 
   \implies G^{1}_{0}\,u^0 -\frac{G^{0}_{0}}{\mu_{D}}\left( G^{1}_{0} \,k^0 +G^{1}_{1}\, k^1\right)=\mu_{N} \,k^1
   \end{align}
 using the relation $ \mu_{N}\,\mu_{D}=G^{1}_{0}G^{0}_{1}-G^{0}_{0}G^{1}_{1}$, we get:
\begin{align}
   u^1&=\frac{G^{1}_{0}}{\mu_{D}}\,k^{0}+\frac{G^{1}_{1}}{\mu_{D}}\,k^1\label{eqn:u1}\\
   u^0&=\frac{G^{0}_{1}}{\mu_{D}}\,k^1 + \frac{G^{0}_{0}}{\mu_{D}}\,k^0\label{eqn:u2}
\end{align}
}
\item {\emph{To obtain explicit expressions for $k^0$ and $k^1$.} To do that, we use the normalization conditions
for the 4-velocity vectors $(u^{\mu}, k^{\lambda})$. 
}
\begin{enumerate}
\item {Using the condition  $g_{\lambda\sigma}\,k^{\lambda}\,u^{\sigma}=1$, we have:
{\small
\begin{widetext}
\begin{align}\label{eqn:firtexp}
    \left( g_{00}G^{0}_{0} +g_{01}G^{1}_{0}\right)(k^0)^2+\left( g_{00}G^{0}_{1} +g_{01}G^{1}_{1}+ g_{01}G^{0}_{0} +g_{11}G^{1}_{0}\right)k^1k^0+ \left( g_{01}G^{0}_{1} +g_{11}G^{0}_{0}\right)(k^1)^2=\mu_{D}
\end{align}
\end{widetext}
}
}
\item {Using the normalization condition for $g_{\lambda \nu}k^{\lambda}\,k^{\nu}=0$, we get:
\begin{equation}
\label{eq:k0k1}
        k^1\,k^0=-\frac{g_{11}}{2g_{01}}(k^1)^2-\frac{g_{00}}{2g_{01}}(k^0)^2 \, .
\end{equation}
Substituting the above expression in Eq.~(\ref{eqn:firtexp}), we get:
\begin{equation}\label{eqn:Isimple}
    (k^0)^2\,P+(k^1)^2\,Q=\mu_{D}
\end{equation}
}
where
{\small
\begin{align}
    P&= \left(g_{00}\,G^{0}_{0} +g_{01}\,G^{1}_{0} \right)\nonumber\\ 
    &-\frac{g_{00}}{2g_{01}}\left( g_{00}\,G^{0}_{1} +g_{01}\,G^{1}_{1}+ g_{01}\,G^{0}_{0} +g_{11}\,G^{1}_{0}\right)\\
   Q&=\left( g_{01}\,G^{0}_{1} +g_{11}\,G^{1}_{1}\right)\nonumber\\
   &-\frac{g_{11}}{2g_{01}}\left(g_{00}\,G^{0}_{1} +g_{01}\,G^{1}_{1}+ g_{01}\,G^{0}_{0} +g_{11}\,G^{1}_{0}\right)
\end{align}
}

\item {Using the normalization condition $g_{\sigma\rho}u^{\sigma}u^{\rho}=1$, we get:
{\small
\begin{widetext}
\begin{align}\label{eqn:secdexp}
 g_{00}\left( \frac{G^{0}_{1}}{\mu_{D}}k^{1}+\frac{G^{0}_{0}}{\mu_{D}}k^{0} \right)^{2}
 +2g_{01}\left( \frac{G^{0}_{1}}{\mu_{D}}k^{1}+\frac{G^{0}_{0}}{\mu_{D}}k^{0} \right)\left( \frac{G^{1}_{0}}{\mu_{D}}k^{0}+\frac{G^{1}_{1}}{\mu_{D}}k^{1} \right)+ g_{11}\left( \frac{G^{1}_{0}}{\mu_{D}}k^{0}+\frac{G^{1}_{1}}{\mu_{D}}k^{1} \right)^{2}  =1
\end{align}
\end{widetext}
}
Substituting Eq.~\eqref{eq:k0k1} in the above expression, we have:
    \begin{equation}\label{eqn:IIsimple}
    (k^0)^2\,R+(k^1)^2\,S=\mu_{D}^2
\end{equation} 
where
{\small
\begin{align}
    S&=\left[ g_{00}\,(G^{0}_{1})^2 +2g_{01}G^{0}_{1}G^{1}_{1} +g_{11}(G^{1}_{1})^2\right]
    \nonumber\\&
    -\frac{g_{11}}{2g_{01}}\left( 2g_{00}G^{0}_{0}G^{0}_{1} +2g_{01}(G^{1}_{0}G^{0}_{1}+G^{0}_{0}G^{1}_{1})+2 g_{11}G^{1}_{0}G^{1}_{1}  \right)
\\
    R&=\left[ g_{00}(G^{0}_{0})^2+2g_{01}G^{1}_{0}G^{0}_{0} +g_{11}(G^{1}_{0})^2 \right] 
    \nonumber\\&
    -\frac{g_{00}}{2g_{01}}\left( 2g_{00}G^{0}_{0}G^{0}_{1} +2g_{01}(G^{1}_{0}G^{0}_{1}+G^{0}_{0}G^{1}_{1})+2 g_{11}G^{1}_{0}G^{1}_{1}  \right)
\end{align}
}
}
\end{enumerate}
{We thus have two equations (\ref{eqn:Isimple}, \ref{eqn:IIsimple}) in two unknown variables $k^0$ and $k^1$, i. e.,
\begin{align}
     (k^0)^2 \, P + (k^1)^2 \, Q=\mu_{D}\\
     (k^0)^2 \, R + (k^1)^2 \, S =\mu_{D}^2
\end{align}
From these we have:
\begin{align}
    (k^0)^2\,(PS-QR)=\mu_{D}(S-\mu_{D}Q )\nonumber\\
    k^0=\pm \,\sqrt{ \frac{\mu_{D}(S-\mu_{D}Q )}{(PS-QR)}}
\\
    (k^1)^2\,(QR-SP)=\mu_{D}(R-\mu_{D}P )\nonumber\\
    k^1=\pm \,\sqrt{ \frac{\mu_{D}(R-\mu_{D}P )}{(QR-SP)}}
\end{align}  
Since we now know $k^0$ and $k^1$, we can obtain 
$u^0$ and $u^1$ from Eqs. (\ref{eqn:u1}) and (\ref{eqn:u2}). Note that if we choose $k^0$ to be positive, $k^1$ can be negative satisfying the condition \eqref{eq:condition01}.
}
\end{STEP}
{For the metric
\begin{align}
ds^2 =&  \left(\frac{\eta}{\eta_{0}}\right)^4 \left[ \left(1-\frac{2 M(\eta)}{r}\right) d\eta^2  -\frac{4 M(\eta)}{r} d \eta\,dr 
\right.\nonumber\\&\left.
~~~~~~~~- \left(1+\frac{2 M(\eta)}{r}\right) dr^2-r^2 d\Omega^2  \right] \, ,
\end{align}
$P, Q, R$ and $S$ are given by:
{\small
 \begin{align}
     P&=\frac{\eta^2\Dot{M}(\eta)+2(6r+\eta)M(\eta)}{r\eta^2M(\eta)}\\
     Q&=-\frac{\eta\Dot{M}(\eta)+2M(\eta)}{r \eta M(\eta)}\\
     S  &= - \left(\frac{\eta_{0}^4 \left[\Dot{M}(\eta)\,\eta + 2M(\eta)\right]}{\eta^7 r^3 M(\eta)}\right)\nonumber\\
     &~~~~\left(24M(\eta)(r - \eta) + 12r^2 + 4\eta \Dot{M}(\eta)(2r - \eta) \right)\\
      R  &=  \left(\frac{\eta_{0}^4 \left[\Dot{M}(\eta)\,\eta^2 + 2M(\eta)(6r+\eta)\right]}{\eta^8 r^3 M(\eta)}\right)\nonumber\\
      &~~~~\left(24M(\eta)(r - \eta) + 12r^2 + 4\eta \Dot{M}(\eta)(2r - \eta) \right)
 \end{align}
 }
 $M(\eta)=M_{\rm I/II}(\eta)$ and $\Dot{M}(\eta)=\frac{dM(\eta)}{d\eta}$.
}

\section{Properties of line-element \eqref{eqn:S2}} 
\label{App:prop}

In this appendix, we list the key properties of the line-element \eqref{eqn:S2}. 
\begin{enumerate}
    \item {\it Apparent horizons} Like line-element \eqref{eqn:S1}, the line-element \eqref{eqn:S2} has two dynamical trapping horizons:
{\small
\begin{align}
    r_{C2} & = \frac{\eta}{2} \label{eqn:DymHr1S2}\\
    r_{H2} & = -M_{II}(\eta)-\frac{\eta}{4}+\frac{\sqrt{16M_{II}(\eta)^2+24\,M_{II}(\eta)\,\eta+\eta^2}}{4} \nonumber
    \label{eqn:DymHr2S2}
\end{align}
}
\item {\it Misner-Sharp-Hernandez energy} is given by: 
{\small
\begin{align}
E_{II}(r,\eta) &=\left(\frac{r}{2}\right)\left(\frac{\eta}{\eta_0}\right)^2 \left[ 1-\left(\frac{1}{\eta^2}\right)\left(4r^2\left(1+\frac{2M_{II}(\eta)}{r}\right)- \right. \right. \nonumber\\ 
& \left. \left.  \qquad \qquad 8\eta\,M_{II}(\eta) -\eta^2\left(1-\frac{2M_{II}(\eta)}{r}\right)   \right)\right]
\end{align}
}
\item {\it Surface gravity:} The dynamical surface gravity for the cosmological horizon is
\begin{equation}
    \kappa_{\rm AH}=\frac{\eta_0^2}{2\eta^4}\left(-12M_{\rm II}(\eta)+\eta\right)
\end{equation}
The dynamical surface gravity at $r= r_{H2}$ is
 \begin{align}
\kappa_{\rm H2}&=\frac{1}{a^2(\eta) r_{H2}^2} 
\left[(M_{II}(\eta) - m)(5\,\eta - 12\,r_{H2})r_{H2} - \right. \nonumber \\
& \left. \qquad M_{II}(\eta)\,\eta^2 -2m\,r_{H2}(\eta - r_{H2}) + r_{H2}^3\right]
 \end{align}
\item {\it Energy density of the dust} is given by:
{\small
\begin{align}\label{eqn:EDB2}
    \mu_{_D}= & \frac{4\eta_0^4}{r^2\eta^{6}}\left[ \sigma_{D}(r,\eta)\, r^{2} + \sigma_{N}(\eta)(r-\eta)\right]\\
     \sigma_{D}(r,\eta) = & 3 + \frac{7\,\eta_{\rm decay}^7m}{\eta^7r};
    \sigma_{N}(\eta) =  m\left[]6 +\frac{\eta_{\rm decay}^7}{\eta^7}\right]
\end{align}
}
where ${\sigma}_{N}(\eta)$ and ${\sigma}_{D}(r,\eta)$ represent the flow of energy along the radial direction and the null direction respectively.

\item {\it Kodama Vector} is the same as in Eq.~\eqref{eq:KodamaVec}. 

\item {It is possible to transform  the solution (\ref{eqn:S2}) to the Branch I solution by setting $ m\rightarrow -m$. To see this, we first 
rewrite the  branch II (\ref{eqn:S2}) line element as: 
\small{
\begin{align}
    ds_{II}^2 &= \frac{\eta^4}{\eta_{0}^4}\left[\left(1-\frac{2\,(-m)}{r}\left[\frac{\eta_{\rm decay}^7}{\eta^7}-1\right]\right)d\eta^2 
      \nonumber\right.\\
    &\left.
       ~~~~~~~~~~~~~-\frac{4\,(-m)}{r}\left[\frac{\eta_{\rm decay}^7}{\eta^7}-1\right]d\eta\,dr
         \nonumber\right.\\
    &\left.
      ~~~~~~ -\left(1+\frac{2\,(-m)}{r}\left[\frac{\eta_{\rm decay}^7}{\eta^7}-1\right]\right)dr^2 -r^2d\Omega \right]
\end{align}}
Setting $m = - M$ in the above line-element corresponds to the Branch I metric with a negative mass $M = -m$. 
Thus, by reversing the mass, the two branches are transformed into each other. We do not consider the second Branch (\ref{eqn:S2}) because it is related to the first Branch (\ref{eqn:S1}) with a negative mass. As we show below, this transformation is unphysical.}

\item {Branch I represents the evolution  $0<\eta < \eta_{\rm decay}$. At $\eta=\eta_{\rm decay}$ the metric becomes conformally Einstein-de-Sitter universe. Further evolution is captured (i.e., $\eta_{\rm decay} < \eta < \infty$) by the branch II (\ref{eqn:S2}) with negative mass$-m$. The branch II solution converges to Sultana-Dyer metric in the limit of $\eta \rightarrow \infty$. Physically, at $\eta=\eta_{\rm decay}$, when the Einstein-de-Sitter space is formed, we have no further information about the formation of the black hole solution in branch II because the black hole is completely evaporated at $\eta = \eta_{\rm decay}$.}
\end{enumerate}

\section{}

As mentioned in the previous appendix, setting $m = - M$ in the branch II line-element corresponds to the Branch I metric with a negative mass $M = -m$. In this appendix, we look at the continuity of the two branches (\ref{eqn:S1}) and (\ref{eqn:S2}).  

\begin{figure*}[!htb]
\begin{minipage}[b]{.5\textwidth}
\includegraphics[width=\columnwidth]{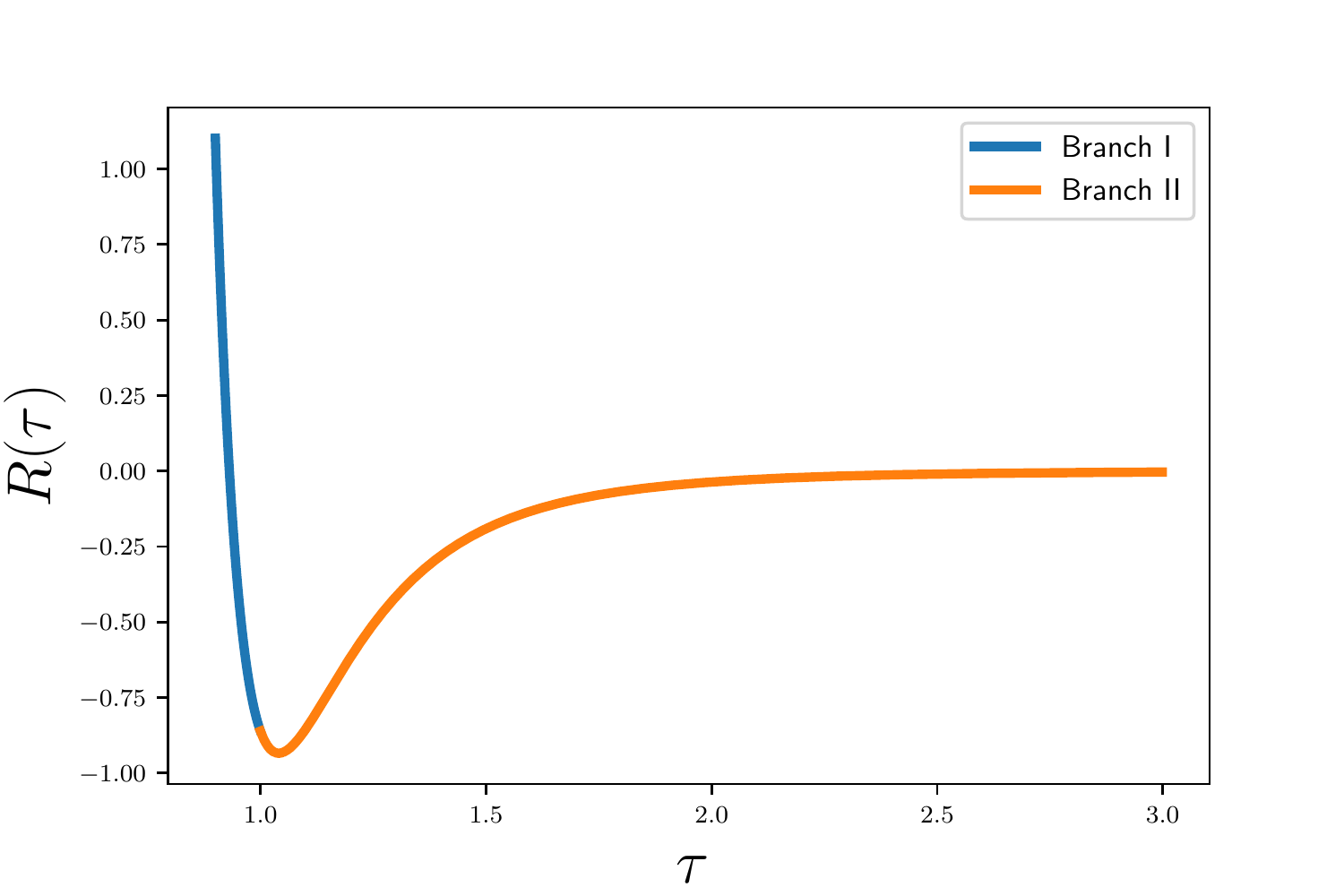}
\end{minipage}\hfill
\begin{minipage}[b]{.5\textwidth}
\includegraphics[width=\columnwidth]{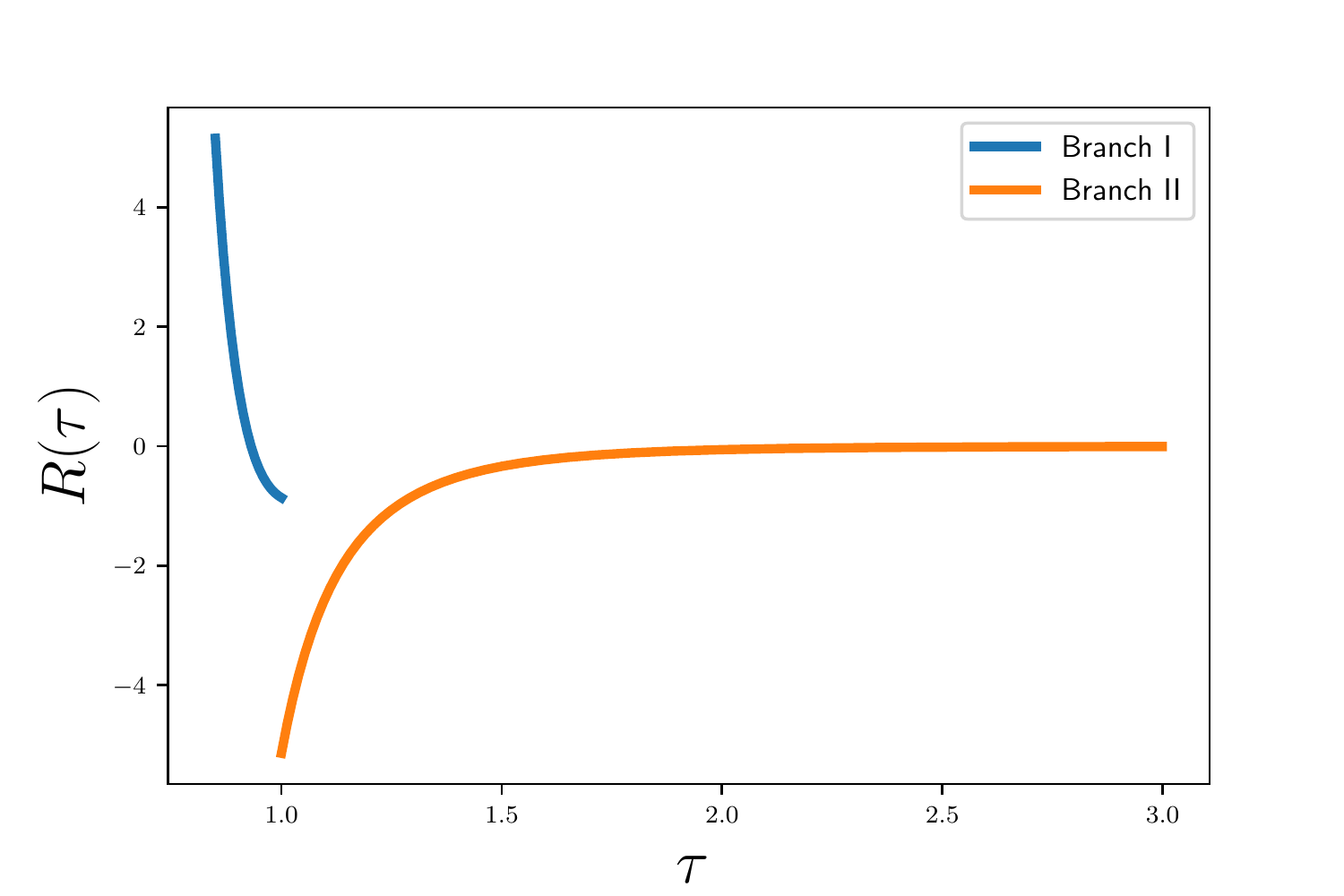}
\end{minipage} 
\caption{\textcolor{blue}{Left}: Plot of the Ricci scalar $(R(\tau))$ vs $\tau$  for the branch I (\ref{Branch1:metr_rep})for $0.95\leq\tau\leq 1,\, m=M_{0}$ and for the branch II (\ref{Branch2:metr_rep(m)}) $1.0\leq\tau\leq3\, ,m=-M_{0}$. \textcolor{blue}{Right}: Plot of the Ricci scalar $(R(\tau))$ vs $\tau$  for the branch I (\ref{Branch1:metr_rep})for $0.85\leq\tau\leq 1,\, m=M_{0}$ and for the branch II (\ref{Branch2:metr_rep(m)}) $1.0\leq\tau\leq3\, ,m=M_{0}$.}
\label{fig:02}
\end{figure*}

To go about that, first, we plot the Ricci scalars 
of the two branches for different mass ranges. Rewriting $\tau = \eta/\eta_{\rm decay}$, the 
line-element and the Ricci scalar 
of the Branch I solution is:
\small{
\begin{align}\label{Branch1:metr_rep}
    ds_{I}^2 &= \frac{\eta_{\rm decay}^4\,\tau^4}{\eta_{0}^4}\left[\left(1-\frac{2m}{r}\left[\frac{1}{\tau^7}-1\right]\right)\eta_{\rm decay}^2\,d\tau^2 
    \nonumber\right.\\
    &\left.
    ~~~~~~~~~~~~~~~~~~~~~-\frac{4m}{r}\left[\frac{1}{\tau^7}-1\right]\eta_{\rm decay}\,d\tau\,dr
    \nonumber\right.\\&\left.~~~~~~~~~~~~~~
    -\left(1+\frac{2m}{r}\left[\frac{1}{\tau^7}-1\right]\right)dr^2-r^2d\Omega \right] \\ 
\label{Branch1:Ricci}
   R_{I} &= -\frac{4 \eta_{0}^{4}\left(6\, \eta_{\rm decay}\, m \tau^{8}-6 m r \tau^{7}+3 r^{2} \tau^{7}+\eta_{\rm decay}\, m \tau-8 m r\right)}{\eta_{\rm decay}^{6} \tau^{13} r^{2}}
\end{align}
}
Similarly, the line-element and the Ricci scalar 
of the Branch II solution is:

\begin{align}\label{Branch2:metr_rep(m)}
    ds_{II}^2 &= \frac{\eta_{\rm decay}^4\,\tau^4}{\eta_{0}^4}\left[\left(1-\frac{2(-m)}{r}\left[\frac{1}{\tau^7}-1\right]\right)\eta_{\rm decay}^2\,d\tau^2 
    \nonumber\right.\\
    &\left.
    ~~~~~~~~~~~~~~~~~~~~~-\frac{4(-m)}{r}\left[\frac{1}{\tau^7}-1\right]\eta_{\rm decay}\,d\tau\,dr
    \nonumber\right.\\&\left.~~~~~~~~~~~~~~
    -\left(1+\frac{2(-m)}{r}\left[\frac{1}{\tau^7}-1\right]\right)dr^2-r^2d\Omega \right] \\
\label{Branch2:Ricci}
     R_{II} &= \frac{4 \eta_{0}^{4}\left(6\, \eta_{\rm decay}\, m \tau^{8}-6 m r \tau^{7}-3 r^{2} \tau^{7}+\eta_{\rm decay}\, m \tau-8 m r\right)}{\eta_{\rm decay}^{6} \tau^{13} r^{2}}
\end{align}
Note that in the case of Branch I, $\tau$ is in the range $[0, 1]$. In the case of Branch II, $\tau$ is in the range $[1, \infty]$. The left panel of Fig. \eqref{fig:02} is the plot of Ricci scalar in both the branches for $0 < \tau < 3$. Ricci scalar of both the branches is continuous at $\tau = 1$, only for negative masses for Branch II. If the mass is positive for both the branches, as shown in the right panel of Fig.\eqref{fig:02}, the Ricci scalar is discontinuous. 
 
To understand that the negative mass ($-m$) corresponds to an unphysical transformation from the Schwarzchild metric, let us repeat the analysis of Sultana and Dyer~\cite{2005-Sultana_Dyer}. 

The Sultana-Dyer metric\cite{2005-Sultana_Dyer} is obtained by conformally transforming Schwarzchild metric. Let us consider the seed metric --- the Schwarzschild vacuum metric with coordinates $\Tilde{t},\Tilde{r}, \Tilde{\theta}$ and $\Tilde{\phi}$:
\begin{align}\label{SchMetric}
    ds^2 = \left(1 -\frac{2\Tilde{m}}{\Tilde{r}}\right)d\Tilde{t}^2 -\frac{d\Tilde{r}^2}{\left(1 -\frac{2\Tilde{m}}{\Tilde{r}}\right)} - \Tilde{r}^2d\Omega^2
\end{align}
which is invariant under the simultaneous transformations of $\Tilde{m} = -\Tilde{m}$ and $\Tilde{r} = -r$. Using the following transformation:
\begin{align}
        \Tilde{r}&=-r\label{SchTrans1}\\
        \Tilde{t} &= t +2\Tilde{m}\, \ln\left(\frac{r}{2\Tilde{m}}+1 \right)\label{SchTrans2}
\end{align}
leads to:
\begin{align}\label{SchMetric2}
    ds^2 &= \left(1 +\frac{2\Tilde{m}}{r}\right)dt^2+\frac{4\Tilde{m}}{r}dt\,dr -\left(1 -\frac{2\Tilde{m}}{r}\right)dr^2 - r^2\,d\Omega^2
\end{align}
Multiplying  (\ref{SchMetric2}) with the conformal factor $t^4$ gives the following metric:
\begin{align}
    ds^2 &= t^4\left[\left(1 +\frac{2\Tilde{m}}{r}\right)dt^2+\frac{4\Tilde{m}}{r}dt\,dr -\left(1 -\frac{2\Tilde{m}}{r}\right)dr^2 - r^2\,d\Omega^2\right] \, .
\end{align}
This metric is identical to the Sultana-Dyer metric ~\cite{2005-Sultana_Dyer}
only if $-\Tilde{m}$ is negative. However, the transformation (\ref{SchTrans1}) is not physically realizable because $r$ is negative definite in the above metric. Hence, the negative mass Sultana-Dyer solutions are not physically realizable. In other words, the two branches correspond to two physically distinct situations. 

\providecommand{\noopsort}[1]{}\providecommand{\singleletter}[1]{#1}%

\end{document}